\documentclass[useAMS,usenatbib]{mn2e}
\usepackage{graphicx}
\usepackage{lscape}

\title[Ross 458C]{The properties of the T8.5p dwarf Ross~458C }
\author[Ben Burningham et al]{Ben Burningham$^{1}$\thanks{E-mail:
    B.Burningham@herts.ac.uk}, S. K. Leggett$^{2}$, D. Homeier$^{3}$, D. Saumon$^{4}$, P.W. Lucas$^{1}$, 
    \newauthor
D. J. Pinfield$^{1}$, C. G. Tinney$^{5}$, F. Allard$^{6}$, M. S. Marley$^{7}$, H.R.A. Jones$^1$, \newauthor
 D. N. Murray$^{1}$, M. Ishii$^{8}$, A. C. Day-Jones$^{9}$, J. Gomes$^{1}$, Z.H. Zhang$^{1}$\\
$^{1}$ Centre for Astrophysics Research, Science and Technology Research Institute, University of Hertfordshire, Hatfield AL10 9AB \\
$^{2}$ Gemini Observatory, 670 N. A'ohoku Place, Hilo, HI 96720, USA \\
$^{3}$ Los Alamos National Laboratory, P.O. Box 1663, MS F663, Los Alamos, NM 87545, USA \\
$^{4}$Institut fur Astrophysik, Georg-August-Universitat, Friedrich-Hund-Platz 1, 37077 Gottingen, Germany \\
$^{5}$ School of Physics, University of New South Wales, 2052. Australia\\ 
$^{6}$C.R.A.L. (UMR 5574 CNRS), Ecole Normale Superieure, 69364 Lyon Cedex 07, France \\
$^{7}$ NASA Ames Research Center, Mail Stop 245-3, Moffett Field, CA 94035, USA \\
 $^{8}$Subaru Telescope, 650 North A'ohoku Place, Hilo, Hi 96720, USA \\
$^{9}$ Universidad de Chile,Camino el Observatorio \# 1515, Santiago, Chile, Casilla 36-D\\
}

\begin{document}
%
%
%
%


\def\aj{\rm{AJ}}                   
\def\araa{\rm{ARA\&A}}             
\def\apj{\rm{ApJ}}                 
\def\apjl{\rm{ApJ}}                
\def\apjs{\rm{ApJS}}               
\def\ao{\rm{Appl.~Opt.}}           
\def\apss{\rm{Ap\&SS}}             
\def\aap{\rm{A\&A}}                
\def\aapr{\rm{A\&A~Rev.}}          
\def\aaps{\rm{A\&AS}}              
\def\azh{\rm{AZh}}                 
\def\baas{\rm{BAAS}}               
\def\jrasc{\rm{JRASC}}             
\def\memras{\rm{MmRAS}}            
\def\mnras{\rm{MNRAS}}             
\def\pra{\rm{Phys.~Rev.~A}}        
\def\prb{\rm{Phys.~Rev.~B}}        
\def\prc{\rm{Phys.~Rev.~C}}        
\def\prd{\rm{Phys.~Rev.~D}}        
\def\pre{\rm{Phys.~Rev.~E}}        
\def\prl{\rm{Phys.~Rev.~Lett.}}    
\def\pasp{\rm{PASP}}               
\def\pasj{\rm{PASJ}}               
\def\qjras{\rm{QJRAS}}             
\def\skytel{\rm{S\&T}}             
\def\solphys{\rm{Sol.~Phys.}}      
\def\sovast{\rm{Soviet~Ast.}}      
\def\ssr{\rm{Space~Sci.~Rev.}}     
\def\zap{\rm{ZAp}}                 
\def\nat{\rm{Nature}}              
\def\iaucirc{\rm{IAU~Circ.}}       
\def\aplett{\rm{Astrophys.~Lett.}} 
\def\apspr{\rm{Astrophys.~Space~Phys.~Res.}}
\def\bain{\rm{Bull.~Astron.~Inst.~Netherlands}} 
\def\fcp{\rm{Fund.~Cosmic~Phys.}}  
\def\gca{\rm{Geochim.~Cosmochim.~Acta}}   
\def\grl{\rm{Geophys.~Res.~Lett.}} 
\def\jcp{\rm{J.~Chem.~Phys.}}      
\def\jgr{\rm{J.~Geophys.~Res.}}    
\def\jqsrt{\rm{J.~Quant.~Spec.~Radiat.~Transf.}}
\def\memsai{\rm{Mem.~Soc.~Astron.~Italiana}}
\def\nphysa{\rm{Nucl.~Phys.~A}}   
\def\physrep{\rm{Phys.~Rep.}}   
\def\physscr{\rm{Phys.~Scr}}   
\def\planss{\rm{Planet.~Space~Sci.}}   
\def\procspie{\rm{Proc.~SPIE}}   

\let\astap=\aap
\let\apjlett=\apjl
\let\apjsupp=\apjs
\let\applopt=\ao

\maketitle

\begin{abstract}
We present near-infrared photometry and spectroscopy, and warm-{\it Spitzer} IRAC photometry of the young very cool T dwarf Ross 458C, which we have typed as T8.5p.  By applying the fiducial age constraints ($\leq 1$Gyr) imposed by the properties of the active M dwarf Ross~458A, we have used these data to determine that Ross 458C has $T_{\rm eff} = 695 \pm 60$K, $\log g = 4.0 - 4.7$ and an inferred mass of 5--20~M$_{Jup}$.  
We have compared fits of the near-infrared spectrum and IRAC photometry to the BT~Settl and Saumon \& Marley model grids, and have found that both sets provide best fits that are consistent with our derived properties, whilst the former provide a marginally closer match to the data for all scenarios explored here. 
The main difference between the model grids arises in the 4.5$\micron$ region, where the BT Settl models are able to better predict the flux through the IRAC filter, suggesting that non-equilibrium effects on the CO-CO$_2$ ratio are important for shaping the mid-infrared spectra of very cool T dwarfs. We have also revisited the issue of dust opacity in the spectra of Ross 458C that was raised by Burgasser et al (2010). We have found that the BT Settl models which also incorporate a condensate cloud model, provide a better match to the near-infrared spectrum of this target than the Saumon \& Marley model with $f_{sed} = 2$, and we briefly discuss the influence of condensate clouds on T dwarf spectra.

\end{abstract}


\section{Introduction}
\label{sec:intro}

Precise determination of the fundamental properties (i.e.
mass, age, metallicity and effective temperature, $T_{\rm eff} $ ) of isolated field brown dwarfs poses some serious challenges. 
These principally arise because (a)
brown dwarfs cool over time, and (b) ages are one of the most difficult
parameters to measure for field stars and brown dwarfs. Together this makes
breaking the degeneracies between age, mass, luminosity and effective
temperature for isolated brown dwarfs extremely difficult.
An object with $T_{\rm eff} = 1000$K, for
example,  might be relatively young and low-mass ( e.g. 500~Myr and
20M$_{Jup}$), or it could equally be older and higher-mass (e.g. 10~Gyr
and 60M$_{Jup}$).  
Furthermore, determination of $T_{\rm eff}$ for such objects is itself
problematic due to the fact that the radii of brown dwarfs must
be estimated using evolutionary models \citep[e.g. ][]{baraffe03,sm08}
which to date have yet to be constrained by direct measurement of
field brown dwarf radii\footnote{Although radii for a handful of brown
  dwarfs in cataclysmic variable  \citep[e.g. ][]{littlefair07} and
  young eclipsing systems \citep[e.g.][]{stassun06} have been measured
these objects are not representative of the field population. }.  
The lack of age constraints for isolated
brown dwarfs makes such estimates particularly uncertain.  
The influence of metallicity on the emergent spectra from brown dwarfs
further complicates matters, and so the calibration of
model spectra using systems with well understood age and metallicity
constraints is key to improving our understanding of sub-stellar
atmospheres.
For these reasons, much effort has been expended identifying and
characterising systems for which dynamical masses may be measured
\citep[e.g. ][ and references therein]{scholz03,dupuy09a,king10} or age
and metallicity fiducially constrained
\citep[e.g. ][]{pinfield06,ben09,adj08,adj10,zz10}.

\citet{goldman10} report the discovery of one such age and
metallicity benchmark system, a candidate T8+ dwarf  identified in the
UKIRT Infrared Deep Sky Survey \citep[UKIDSS; ][]{ukidss} that has
been shown to be a
common proper motion companion to the M0.5+M7 binary system
Ross~458AB.  
Using gyrochronology arguments and the age-activity relations of
\citet{west08}, \citet{goldman10} constrain the age of Ross~458AB to
less than 1Gyr.  
They also highlight its possible membership of the Hyades moving group,
although conflicting radial velocity measurements by \citet{hawley97}
and \citet{nidever02} leave this in doubt. 
However, given the spread in age and metallicity seen for members of
the Hyades moving group, it is not immediately clear that such membership would
permit any tighter constraints to be placed on the properties of the
Ross~458 system than are possible otherwise. 
\citet{burgasser2010} used the $M_K$/metallicity and $V-K$/metallicity
relations of \citet{johnson2009} and \citet{schlaufman2010} to
estimate metallicities of [M/H]$ = +0.20 \pm 0.05$ and $0.31 \pm 0.05$
for the Ross~458 system.
The young ($\leq$ 1~Gyr) age of the system, though, means that it is not
immediately clear that such relations can be applied with
confidence. We return to this point at the end of
Section~\ref{subsec:models}. 

The first estimate of Ross 458C's spectral type was made through deep
methane imaging by \citet{goldman10}.  
 It was subsequently and independently identified as part of an
 ongoing large program to 
 identify the coolest T dwarfs in the UKIDSS Large Area Survey (LAS)
 \citep[e.g.][]{lod07,pinfield08,ben08,ben09,ben10a}.
As part of that program we have obtained near-infrared
spectroscopy, as well as additional near-infrared and warm-{\it
  Spitzer} photometry which we use here to estimate the  properties for
this object.  

Recently, \citet{burgasser2010} has presented spectroscopy of
Ross~458C deriving a spectral type of T8. They also compared their
spectroscopy to a number of \citet{saumon06,saumon07} model spectra to
derive properties. They found that model spectra with significant
condensate opacity provided a better match to their data than
cloudless models. 
In Section~\ref{subsec:models} we compare our data to models from the
Saumon \& Marley grid, and also to the latest BT Settl model grid, which also includes a condensate cloud model \citep{allard2010}.

\section{New observations of Ross~458C}
\label{sec:newobs}

\subsection{Near-infrared photometry}
\label{subsec:photo}

Near-infrared follow-up photometry was obtained using the Wide Field
CAMera \citep[WFCAM; ][]{wfcam} on UKIRT on the night of 17$^{th}$
June 2009, and 
the data were processed using the WFCAM science pipeline by the
Cambridge Astronomical Surveys Unit (CASU) \citep{irwin04}, and
archived at the WFCAM Science Archive \citep[WSA; ][]{wsa}.
Observations consisted of a three point jitter pattern in the $Y$ and
$J$ bands, and five point jitter patterns in the $H$ and $K$ bands
repeated twice. All data were acquired with 2x2 microstepping and individual exposures of
10 seconds resulting in total integration times of 120 seconds in $Y$
and $J$ and 400 seconds in $H$ and $K$. 
The WFCAM filters are on the Mauna Kea Observatories (MKO) photometric
system \citep{mko}.

The resulting photometry is given in Table~\ref{tab:photo}. It should
be noted that the follow-up WFCAM data are generally in good agreement
with the survey data, with the exception of the $Y$~band, for which there
is a 9$\sigma$ disagreement. 
The origin of this discrepancy is not clear.

\subsection{Warm-{\it Spitzer} photometry}

Data were obtained for Ross 458C on 13$^{th}$~March 2010, via the Cycle 6
GO program 60093. 
Individual frame times were 30 seconds repeated three times, with a
16-position spiral dither pattern, for a total integration time of 24
minutes in each of the [3.6] and [4.5] bands. 
The post-basic-calibrated-data (pbcd) mosaics generated by version
18.14.0 of the Spitzer pipeline were used to to obtain aperture
photometry.  
The photometry was derived using a 7$\arcsec$ aperture, and the
aperture correction was taken from the IRAC
handbook\footnote{http://ssc.spitzer.caltech.edu/irac/dh/}.  
The error is estimated by the larger of either the variation with sky
aperture, or the error implied by the uncertainty images. This error
is small (Table~\ref{tab:photo}) and is dominated by the 3\% error
that  should be added in quadrature to the quoted random errors to account for
systematic effects due to calibration uncertainties and pipeline
dependencies.

\begin{table*}
\begin{tabular}{| c c c c c|}
$z'_{SDSS}$ & $Y_{UKIDSS}$ & $J_{UKIDSS}$ & $H_{UKIDSS}$ & $K_{UKIDSS}$ \\
\hline
$20.24 \pm 0.19$ & $17.72 \pm 0.02$ & $16.69 \pm 0.01$ & $17.01 \pm 0.04$ & $16.90 \pm 0.06$ \\
\hline
\end{tabular}
\begin{tabular}{| c c c c |}
$Y_{WFCAM}$ &  $J_{WFCAM}$ & $H_{WFCAM}$ & $K_{WFCAM}$ \\
\hline
$17.54 \pm 0.02$ & $16.71 \pm 0.01$ & $17.07 \pm 0.03$ & $16.96 \pm 0.03$ \\
\hline
\end{tabular}
\begin{tabular}{| c c |}
[3.5] &  [4.5] \\
\hline
$15.28 \pm 0.01$ & $13.77 \pm 0.01$  \\
\hline
\end{tabular}

\begin{tabular}{| c c c c c c|}
$z'_{SDSS}-J_{WFCAM}$ & $Y_{WFCAM}$ - $J_{WFCAM}$ & $J_{WFCAM} - H_{WFCAM}$ & $H_{WFCAM}- K_{WFCAM}$ & $H_{WFCAM} - {\rm [4.5]}$ & [3.5]--[4.5] \\
\hline
$3.53 \pm 0.19$ & $0.83 \pm 0.02$ & $-0.36 \pm 0.03$ & $0.11 \pm 0.04$ & $3.30 \pm 0.03$ & $1.15 \pm 0.02$\\
\hline
\end{tabular}

\caption{Summary of photometry for Ross 458C.  The subscripts on the column headings indicated the source of the data, where: ``UKIDSS" indicates UKIDSS survey data; ``SDSS" indicates SDSS DR7 survey photometry; ``WFCAM" indicates WFCAM follow-up photometry.
\label{tab:photo}}

\end{table*}

\subsection{Near-infrared spectroscopy}
\label{subsec:spectra}

We used $JH$ and $HK$ grisms in the InfraRed Camera and Spectrograph
\citep[IRCS;][]{IRCS2000} on the 
Subaru telescope on Mauna Kea to obtain a R$\sim 100$ $JH$ and $HK$ spectra for
Ross~458C on 7$^{th}$ May 2009 and 31$^{st}$ December 2009 respectively.
The observations were made up of a set of four 300s sub-exposures for
the $JH$ spectrum and six 240s sub-exposures (obtained
in an ABBA jitter pattern to facilitate background subtraction),
with a slit width of 0.6 arcsec delivering a resolution of R$\sim$100.  
In addition we obtained a deep $H$ band spectrum using the Near InfraRed
Imager and Spectrometer \citep[NIRI;][]{hodapp03} on the Gemini North
Telescope\footnote{under program GN-2009B-Q-62}. The NIRI observations
were made up of twelve 300s sub-exposures observed
in an ABBA jitter, with a 1 arcsecond slit delivering a resolution of
R$\sim$500.

The spectra were extracted using standard IRAF
packages. The AB pairs were subtracted using generic IRAF tools,
and median stacked.  
In the case of IRCS, the data were found to be sufficiently uniform in
the spatial axis for flat-fielding to be neglected.
We used a comparison argon arc frame to obtain the dispersion
solution, which was then applied to the pixel coordinates in the
dispersion direction on the images. 
The resulting wavelength-calibrated subtracted pairs had a low-level
of residual sky emission removed by fitting and subtracting this
emission with a set of polynomial functions fit to each pixel row
perpendicular to the dispersion direction, and considering pixel data
on either side of the target spectrum only. 
The spectra were then extracted using a linear aperture, and cosmic
rays and bad pixels removed using a sigma-clipping algorithm.

Telluric correction was achieved by dividing each extracted target
spectrum by that of an F5 star, observed just
before or after the target and at a similar airmass. 
Prior to division, hydrogen lines were removed from the standard star
spectrum by interpolating the stellar
continuum.
Relative flux calibration was then achieved by multiplying through by a
blackbody spectrum of the appropriate $T_{\rm eff}$.

The IRCS $JH$ and $HK$ spectra were joined by scaling them to match at
the $H$ band peak (at 1.58$\micron$). The higher signal-to-noise NIRI
$H$ band spectrum was then used to replace the $H$ band region of the
IRCS spectra, again by scaling it to match the IRCS spectrum at
1.58$\micron$.
We note here that the NIRI spectrum closely traced the shape of the
noisier IRCS spectra in this region.

\subsection{Spectral type}
\label{subsec:sptype}

\begin{figure*}
\includegraphics[height=500pt, angle=90]{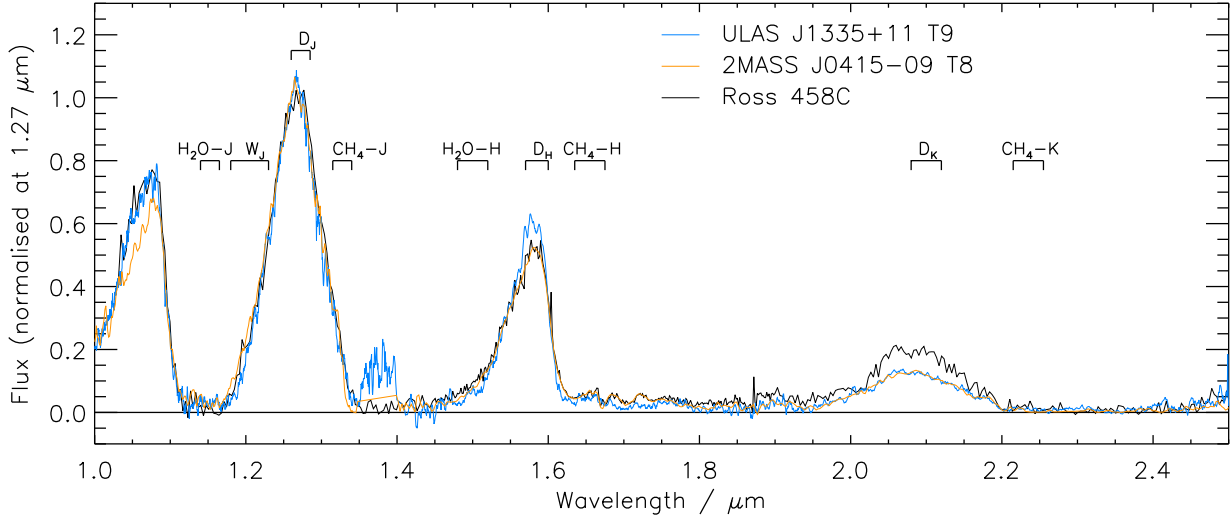}
\caption{The IRCS+NIRI $JHK$ spectrum for Ross~458C plotted with T8 and
  T9 spectral standards 2MASS~J04151954--093506.6 and
  ULAS~JJ133553.45+113005.2. Overlaid are the spectral ranges for the
  numerators (indicated by the index for which they apply) and denominators (D; subscript indicates the index set) of the spectral typing indices used in Table 1
   and outlined in \citet{burgasser06} and \citet{ben08}.}
\label{fig:JHKspec}
\end{figure*}

We have assigned a spectral type for Ross~458C following the scheme
laid out by \citet{burgasser06} and extended by \citet{ben08}.
Figure~\ref{fig:JHKspec} shows our combined IRCS+NIRI $JHK$
spectrum for Ross~458C along with spectra for the T8 and T9 templates
2MASS~J04151954--093506.6 and ULAS~J133553.45+113005.2. 
Whilst the spectrum of Ross~458C lies roughly between the T8 and T9
spectra in the $J$ band, it appears to be earlier in type in the
H$_2$O--$H$ region, and there is considerable excess flux in the $K$
band. 

The spectral flux ratios for Ross~458C are given in
Table~\ref{tab:indices}. Their values reflect the qualitative
comparison with the spectral templates described above. 
The mismatch in inferred spectral type between the three indices that are not
degenerate ($W_J$, H$_2$O--$H$, CH$_4$--$K$) makes assigning a spectral
type for this object problematic.
The remaining index values are all consistent with a type
later than T7, in disagreement with the H$_2$O--$H$ and CH$_4$--$K$
indices but in agreement with the $W_J$ index.
We thus allow the $W_J$ index to dominate the classification, and
assign a type of T8.5p ($\pm 0.5$), reflecting the close agreement between
the T8 and T9 spectra in the $J$ band, and highlighting the
peculiarity of the spectrum elsewhere.

This spectral type is consistent with that estimated for Ross~458C
from methane imaging  \citep[T8.9, ][]{goldman10}, and also that found
by \citet{burgasser2010}. 
However, it disagrees with the type of T7
found by \citet{scholz10b} using a combination of $JHK$ colour-colour
plots and absolute magnitude arguments. 
This is driven by the weight \citet{scholz10b} places on the estimates
from $J-H$/ $J-K$ colour-colour plots, which suggest an early
type of T4.0--T6.5. 
Since the colours of Ross~458C are fairly typical of other T8+ dwarfs
discovered to date, it would appear that such plots have little
utility in spectral typing the latest type T dwarfs.

The disagreement between the $W_J$ index and the H$_2$O--$H$ index
that is seen in Ross~458C is
broadly consistent with the unusual spectral morphology referred to as
``H$_2$O--$H$-early'' peculiarity by \citet{ben10b}.
In addition to Ross~458C, two other $\geq$T7 dwarfs have been identified
that display this peculiarity: the T8p dwarf ULAS~J101721.40+011817.9
\citep{ben08} and the T7p dwarf Gl~229B \citep{nakajima95,burgasser06}.
Several earlier type T~dwarfs have also been identified with this
peculiarity and are described in \citet{ben10b}.

It is interesting to note that both Gl~229B and ULAS~J1017+0118 have
inferred ages similar to that for Ross~458C, the former implied by
the activity of its M0.5 primary star \citep[e.g.][]{west08} and the
latter by comparison of near- and mid-infrared photometry to model
predictions \citep{sandy10}. 
However, interpretation of the H$_2$O--$H$ early peculiarity as being
directly 
connected to the relative youth of these systems is probably premature,
particularly in light of the recent observation that amongst the
ealier-type objects that display this morphology described in
\citet{ben10b}, one has emerged to exhibit halo kinematics (Murray
et al. in prep). 


\begin{table}\renewcommand{\arraystretch}{3}\addtolength{\tabcolsep}{-1pt}
\begin{tabular}{c c c c }
  \hline
 {\bf Index} & {\bf Ratio} & {\bf Value} & {\bf Type} \\
\hline
H$_2$O-J & $\frac{\int^{1.165}_{1.14} f(\lambda)d\lambda}{\int^{1.285}_{1.26}f(\lambda)d\lambda }$ & $0.007 \pm 0.004$  & $\geq$T8 \\[+1mm]
CH$_4$-J & $\frac{\int^{1.34}_{1.315} f(\lambda)d\lambda}{\int^{1.285}_{1.26}f(\lambda)d\lambda }$ &  $0.20 \pm 0.01$ & $\geq$T8 \\
$W_J$ & $\frac{\int^{1.23}_{1.18} f(\lambda)d\lambda}{2\int^{1.285}_{1.26}f(\lambda)d\lambda }$   &  $0.27 \pm 0.01$  & T8/9 \\
H$_2$O-$H$ & $\frac{\int^{1.52}_{1.48} f(\lambda)d\lambda}{\int^{1.60}_{1.56}f(\lambda)d\lambda }$ & $0.22 \pm 0.01$  & T7 \\
CH$_4$-$H$ & $\frac{\int^{1.675}_{1.635} f(\lambda)d\lambda}{\int^{1.60}_{1.56}f(\lambda)d\lambda }$ & $0.11 \pm 0.01$ & $\geq$T8 \\
NH$_3$-$H$ &  $\frac{\int^{1.56}_{1.53} f(\lambda)d\lambda}{\int^{1.60}_{1.57}f(\lambda)d\lambda }$ & $0.71 \pm 0.01$ & ... \\
CH$_4$-K &  $\frac{\int^{2.255}_{2.215} f(\lambda)d\lambda}{\int^{2.12}_{2.08}f(\lambda)d\lambda }$ & $0.20 \pm 0.01$ & T5 \\
\hline
\end{tabular} 
\caption{The spectral flux ratios for Ross~458C. The locations of the numerators and denominators are indicated on Figure~\ref{fig:JHKspec}.The NH$_3$
index is not used for assigning a type \citep[see ][ and Burningham et
al 2010 for a discussion of this]{ben08}, but is included for
completeness and to permit future comparison with other late T~dwarfs. }
\label{tab:indices}
\end{table}

\section{The properties of Ross 458C}
\label{sec:properties}

\subsection{Luminosity and effective temperature}
\label{subsec:teff}

We estimate the bolometric flux ($F_{bol}$) of Ross 458C
following the method outlined in \citet{ben09},
by combining our $JHK$ spectra (flux calibrated by our WFCAM
follow-up photometry) with model spectra (to allow us
to estimate the flux contributions from regions outside our
near-infrared spectral coverage).
We have scaled the $\lambda < 1.05 \micron$ region of the models to
match the flux level in our $JHK$ spectrum, whilst we have used the
warm-{\it Spitzer} channel 1 and 2 photometry to scale the 2.5--4.0 $\micron$
and $\lambda > 4.0 \micron$ regions respectively. To provide some indication
of the systematic effects of our choice of atmospheric models we have
estimated the bolometric flux using both BT~Settl models
\citep{settl08} and those of \citet{saumon06,saumon07}.

We initially selected BT~Settl model spectra covering the $T_{\rm eff} = 500-750$K range,
$\log g = 4.0-5.0$ and [M/H]$= 0.0$ and +0.3. 
This range of parameters was selected to ensure that we include all
likely possibilities for such a young and late-type T dwarf, and
allowing for the potentially high metallicity suggested by \citet{burgasser2010}. 
We then took the median as our value for $F_{bol}$, and the scatter in
values as an estimate of the systematic uncertainty associated with
our use of theoretical extensions scaled to photometry, amounting to
approximately 11\%. 
The uncertainties in the photometry used to scale the model spectra
contribute approximately 4\% to the flux error budget. 
We thus estimate for Ross~458C that $F_{bol} = 6.40 \pm 0.80 \times
10^{-16}$Wm$^{-2}$, of which approximately 60\% is emitted longward of
2.4$\micron$. 
Using the well determined distance to the primary star, and
incorporating the uncertainty therein, we estimate the total
luminosity of Ross~458C to be $2.59 \pm 0.34 \times 10^{-6} L_{\sun}$.



We used the \citet{saumon06,saumon07}
models to fill in the missing spectral regions in an identical manner
to that described above for the BT~Settle model grid. 
We selected cloudless models across the same temperature, gravity and
metallicity range as used for the BT~Settl bolometric correction, and for
simplicity at this stage we have considered only a single eddy
diffusion coefficient (which parameterises non-equilibrium effects due
to turbulent mixing - see below) of $\log K_{zz} = 4$. 
This value has been selected as simply a typical one for late T dwarfs, and does represent a firm estimate for the value of this parameter for Ross~458C, and selecting a different value does not affect our flux estimate, since we scale our model spectra by the observed photometry. In Section~\ref{subsec:models} we find the best fitting value of $\log K_{zz}$.

We find that this choice of models for our bolometric correction
suggests a value of $F_{bol} = 6.06 \pm 0.42 \times
10^{-16}$~Wm$^{-2}$, with a 7\% contribution to the uncertainty
arising from the scatter in the model extensions.  We find that essentially the same contribution to the correction arises from the
region longward of 2.4$\micron$. 
Our estimate for the total luminosity of Ross~458C using the Saumon \&
Marley model set is thus $2.45 \pm 0.20 \times 10^{-6} 
L_{\sun}$, which is consistent with the estimate derived using the
BT~Settl model grid.   
Both estimates for $L_{bol}$ are substantially brighter than has been
found for the similarly typed late-T dwarf Wolf 940B
\citep{ben09,sandy10} for which $9.79 \times 10^{-7} L_{\sun}$ has
been estimated \citep{sandy10}.


To determine $T_{\rm eff}$ for Ross 458C we have used the constraints
placed on its age by \citet{goldman10}  to estimate a likely range of
radii.  
As with the bolometric correction, we have used two sets of
evolutionary models to provide some indication of the systematic
effects involved in this process. 
Using predictions from the COND \citep{baraffe03} evolutionary
models, for the $0.1 \leq {\rm age} \leq 1$Gyr range, we find that our
derived luminosity estimates are consistent with a radius in the
0.101--0.120~R$_{\sun}$ interval (and $\log g = 4.7$ and 4.0
respectively).  
This results in an inferred $T_{\rm eff}$ range of  $730 \pm 25$K to
$670 \pm 20$K for our BT~Settl based bolometric correction, and a
range of $720 \pm 15$ to $660 \pm 15$ for our Saumon \& Marley
corrected luminosity.

Using the evolutionary tracks of \citet{sm08} suggests a slightly
larger radius than the COND models, with a radius in the range
0.103--0.124~R$_{\sun}$ (and $\log g = 4.7$ and 4.0 respectively),
resulting in lower inferred values for $T_{\rm  eff}$, with ranges of
$720 \pm 25$K to $660 \pm  20$K and $710 \pm 15$K to $650 \pm 15$K
for our BT~Settl and Saumon \& Marley corrected luminosities
respectively.  

Although the BT~Settl bolometric correction suggests a marginally
higher $T_{\rm eff}$ range, the two estimates are consistent and we
adopt $T_{\rm eff} = 695 \pm 60$K as our final estimate, with $\log g
= 4.0-4.7$. 
For our age constraint of 0.1--1Gyr this corresponds to a mass range
of roughly 5--20~M$_{J}$ according to both sets of evolutionary
models.

\subsection{Model comparison}
\label{subsec:models}

\begin{figure*}
\includegraphics[height=500pt, angle=90]{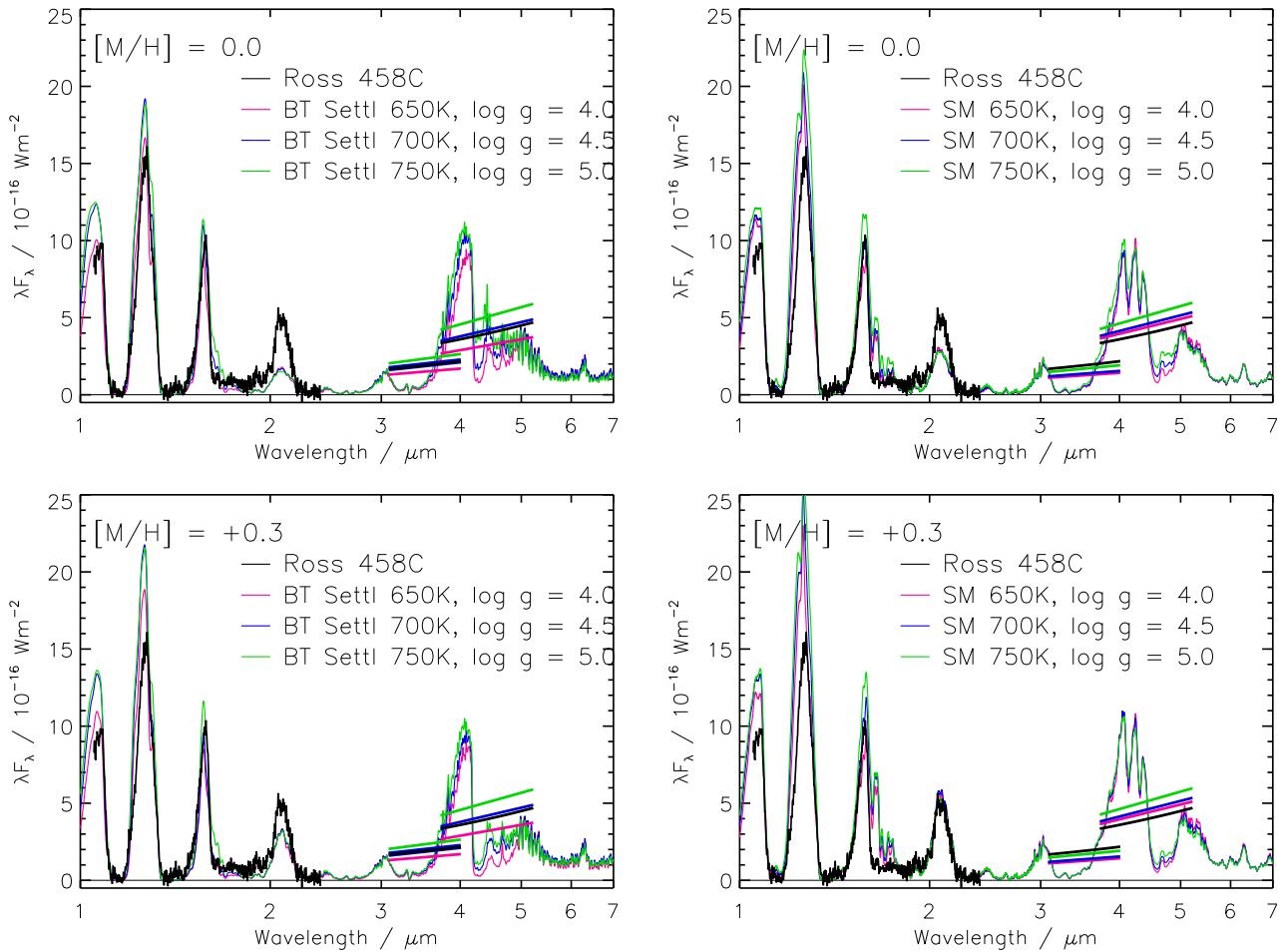}
\caption{The $JHK$ spectrum of Ross~458C and mean fluxes inferred from
  the warm-{\it Spitzer} photometry compared to model spectra that
  straddle the properties estimated in
  Section~\ref{subsec:teff}.
In the case of the Saumon \& Marley models, we found that the model
grid for which $\log K_{zz} = 6$, provided the best match to the data,
and this is what has been plotted here.
The straight coloured lines indicate the mean fluxes of model spectra
in the Spitzer photometric bands and plotted to allow comparison with
the mean flux from the target (straight black lines).
}
\label{fig:speccomp}
\end{figure*}

In Figure~\ref{fig:speccomp}, we have compared model spectra for a range of
parameters that straddle our $T_{\rm eff}$ and gravity estimates for
Ross~458C with the observed spectrum and the mean fluxes inferred for
the warm-{\it Spitzer} channel 1 and 2 photometry.
On the whole the BT~Settl models appear to provide the best fits in
the near-infrared although, as has been seen before \citep{ben09},
their $K$~band flux is substantially underestimated.  
Of note also is the difference between the two model sets in the
1.57-1.60$\micron$ region, where missing methane opacities are thought
to be a significant factor in the generally poor fit of model spectra
to data in this region.  
Both model sets base their methane opacities on the same incomplete
line lists, however the BT~Settl model grid use a statistical estimate of the contributions from hot vibrational bands in the $H$ and $K$ bands, based on the results of \citet{bcjatmos}, and this may account for the differences seen in the $H$~band absorption features.

The most notable differences between the two model sets, however, lie
in the 3.5--5$\micron$ region, where the BT Settl models are able to
match both the [3.5] and [4.5] fluxes well. 
The Saumon \& Marley models, on the other hand, tend to somewhat
underestimate the flux in the former and overestimate the flux in the
latter.
Both model sets predict similar absorption strengths for the CO fundamental
band at 4.55$\micron$, but a significant difference in the integrated [4.5] fluxes
arises from a much stronger CO$_2$ band centered at 4.3$\micron$ in the BT~Settl
models than is seen in the Saumon \& Marley models. Both sets of models
use similar CO$_2$ opacities and the same CNO abundances, but the BT~Settl
consider chemical non-equilibrium effects on the CO$_2$ to CO mixing ratio and thus could be expected have higher carbon dioxide abundances in the upper atmosphere.

\begin{figure}
\includegraphics[height=250pt, angle=90]{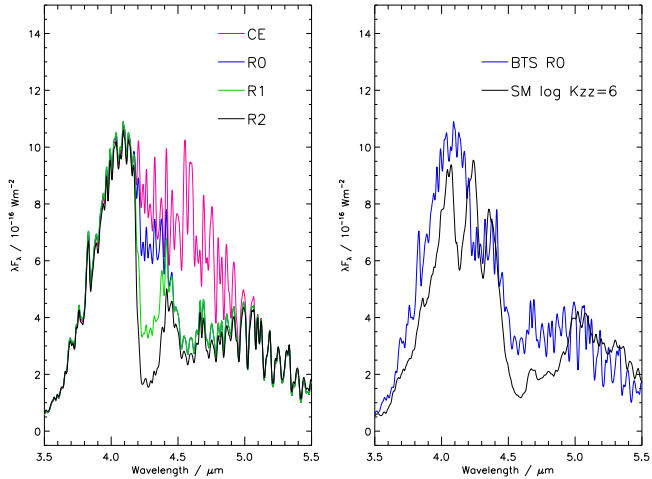}
\caption{{\bf Left:} A comparison of different treatments of carbon chemistry in the BT Settl model grid for a Solar metallicity, $T_{\rm eff} = 700$K, $\log g = 4.5$ atmosphere where: CE = carbon species in full chemical equilibrium; R0 = equal reduction rate for CO$_2$ and CO; R1 = includes the reaction of CO with OH for the conversion between CO and CO$_2$; R2 = as for R1 but also includes the reaction of CO with H$_2$O. {\bf Right:} A comparison of the R0 case with the Saumon \& Marley model for the same atmospheric parameters and $\log K_{zz} = 6$.
}
\label{fig:co2comp}
\end{figure}

To examine this possibility in more detail we have recalculated the BT Settl model spectrum for the $T_{\rm eff} = 700$K, $\log g = 4.5$, Solar metallicity case under several different assumptions about CO and CO$_2$ chemistry.
The left side of Figure~\ref{fig:co2comp} compares the 3.5--5.5$\micron$ region for the BT Settl models under the following cases: CE - carbon species in full chemical equilibrium; R0 - equal reduction rate for CO$_2$ and CO; R1 - includes the reaction of CO with OH for the conversion between CO and CO$_2$, as used by \citet{visscher2010} for the Jovian atmosphere; R2 - this is the standard BT Settl treatment and is as for R1 but also includes the reaction of CO with H$_2$O. 
The R0 case is roughly equivalent to the treatment in the Saumon \& Marley models, and the right side of Figure~\ref{fig:co2comp} shows that this case compares much more closely than the two CO$_2$ non-equilibrium cases. This supports the assertion that the principal cause of the difference between the two model groups in this region is the non-equilibrium treatment for CO$_2$ chemistry used by the BT Settl models.

We note that the default BT Settl (R2) case produces a between five and six times higher CO$_2$ abundance than with the single reaction for converting CO to CO$_2$, and 30--40 times higher CO$_2$ abundance than in the case where both CO and CO$_2$ are quenched at the same level (as in the Saumon \& Marley models).  The increased abundance of CO$_2$ does not impact the CO-CH$_4$ chemistry significantly since CO remains over 20 times more abundant than CO$_2$ even in the default BT Settl (R2) model, and CH$_4$ number density is similarly unaffected by any of the chemistry models since it always remains two orders of magnitude more abundant than CO.

It is clear then that it is appropriate to attribute the difference between the models in the 4.3$\micron$ region to the non-equilibrium treatment of CO$_2$ in the BT Settl models, and that the strength of the CO$_2$ absorption is also strongly dependent on which reaction paths are considered. 
The better fit of the BT Settl models to the observed warm-{\it Spitzer} photometry suggests that such effects are important for determining the emergent spectra of very cool T dwarfs.
The non-equilibrium treatment for CO$_2$ chemistry is further supported by the recent identification of strong 4.3$\micron$ CO$_2$ absorption in {\it AKARI} / IRC spectroscopy of the T8 dwarf 2MASS~J04151954-0935066 \citep{yamamura2010}, although it is not yet established if both reaction pathways considered in the BT Settl models are at work in such environments.

To quantitatively assess the best fitting model
 we have employed the goodness-of-fit statistic, $G$,  of \citet{cushing08} which is defined for a given model, {\it k} as

\begin{equation}
G_k  = \sum_{i=1}^{n} w_i  \left( \frac{f_i - C_k F_{k,i}}{\sigma_i} \right )^2
\label{eqn:Gk}
\end{equation}

where $w$ is the weight to be assigned to wavelength interval, $i$,
$f$ is the observed flux with uncertainty $\sigma$,  $F$ is the model
flux and $C$ is a scaling factor equal to $R^2 / D^2$ for a source of
radius $R$ at distance $D$. 
We take the value of the weight $w$ as the
width in wavelength space of each point on the spectrum.  
In all cases we convolved the model spectra with the instrumental
profile of the observed spectrum of Ross~458C such that the
resolutions are equivalent.
In applying this statistic we have treated the total flux captured
through each warm-{\it Spitzer} filter as a single point, weighted by
the width of the filter. 
Uncertainties in the value of $G$ have been calculated by performing 10000 Monte Carlo simulations for each case, randomly offsetting the target spectrum according to the measured uncertainties.

\begin{figure}
\includegraphics[height=250pt, angle=90]{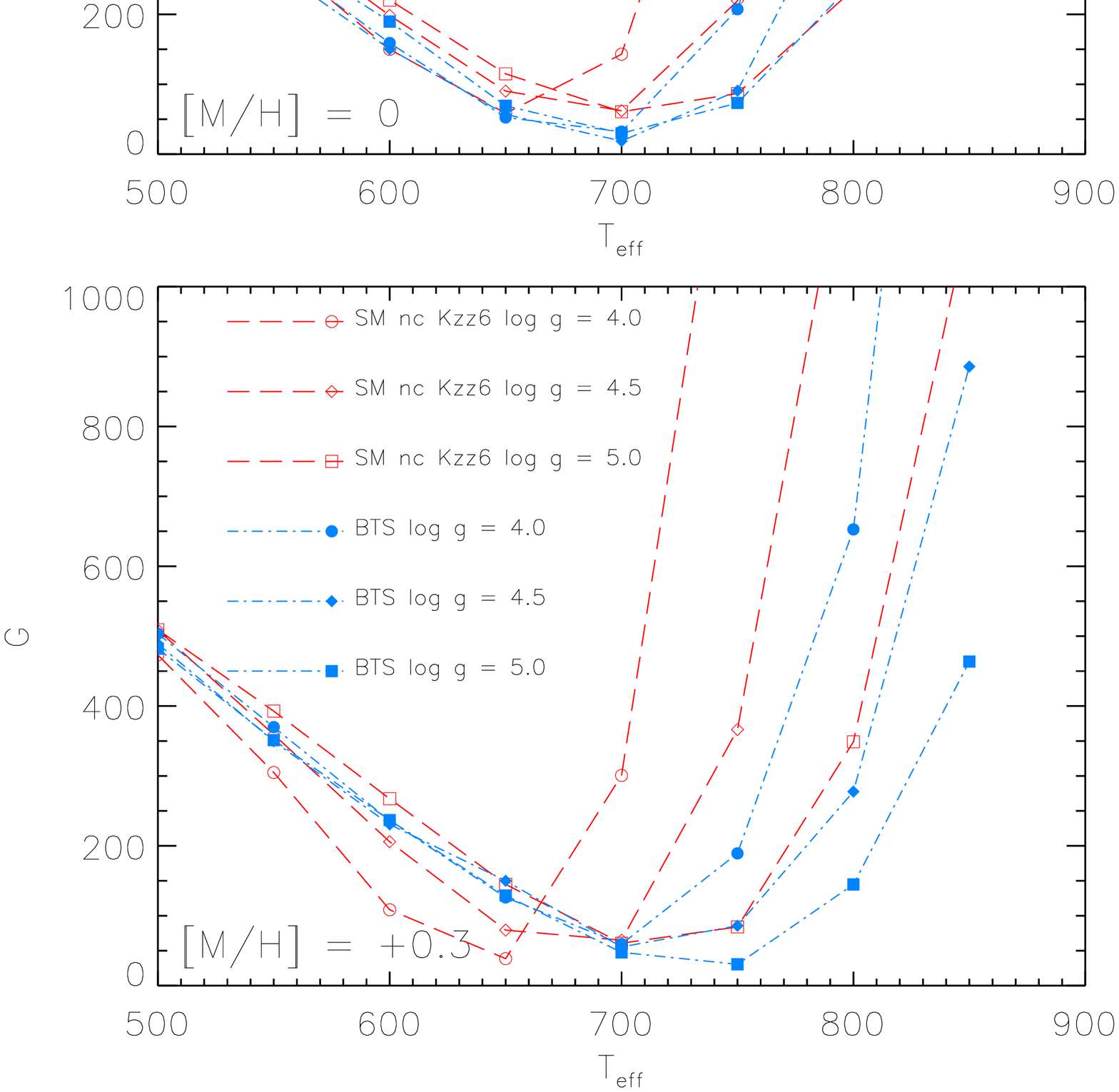}
\caption{The goodness-of-fit statistic, $G$, plotted for various model
  spectra with the value of the scaling factor $C$ fixed by
  evolutionary considerations and the measured parallax to Ross~458AB.
For simplicity we only plot the best fitting case of $\log K_{zz} = 6$
for the Saumon \& Marley model grid.}
\label{fig:fixedCk}
\end{figure}

\begin{figure}
\includegraphics[height=250pt, angle=90]{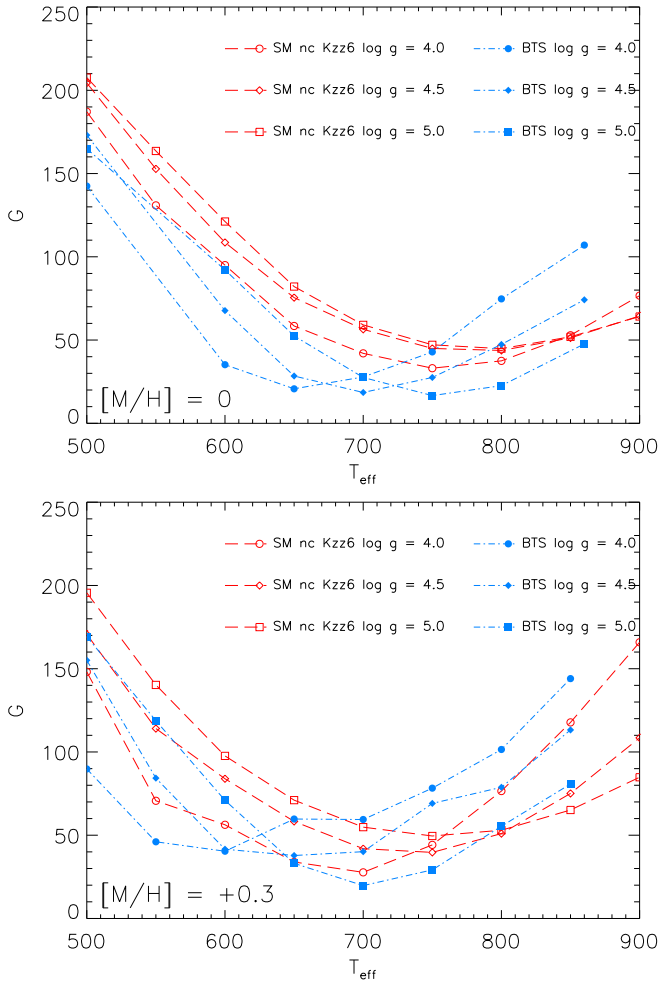}
\caption{The goodness-of-fit statistic, $G$, plotted for various model
  spectra with the value of the scaling factor $C$ evaluated such that
  $G$ is minimised in each case \citep[see ][]{cushing08}.For
  simplicity we only plot the best fitting case of $\log K_{zz} = 6$ for
  the Saumon \& Marley model grid. } 
\label{fig:freeCk}
\end{figure}

In Figure~\ref{fig:fixedCk} we have plotted the  values for $G$ for
various $T_{\rm eff}$ and $\log g$ for both sets of models.
 We have computed $C$ using the known parallax of Ross~458AB and radii
consistent with the values of $\log g$ as implied by both sets of
evolutionary models. 
We have also calculated $G$ for the case where the scaling factor $C$
is unconstrained by parallax and evolutionary arguments to determine
the $T_{\rm eff}$ estimate that would be found from pure spectral
fitting in the case of the two sets of models, and the results are plotted in Figure~\ref{fig:freeCk}.
We have summarised the best fitting models for each case in Table~\ref{tab:bestfits}.

\begin{table*}
\begin{tabular}{c c c c c c c}
\hline
{\bf Case} & \multicolumn{3}{|c|}{\bf BT~Settl best fit} &  \multicolumn{3}{|c|}{\bf SM best fit} \\
 & $T_{\rm eff}$ & $\log g$ & $G$ & $T_{\rm eff}$ & $\log g$ & $G$ \\
\hline
${\rm [M/H]} = 0.0$, fixed $C_k$ & 700~K & 4.5 & $18.7 \pm 0.8$ & 650~K & 4.0 & $58.5 \pm 6.8$ \\
$ {\rm [M/H]} = +0.3$, fixed $C_k$ & 750~K & 5.0 &  $30.5 \pm 2.7$ & 650~K & 4.0 & $38.6 \pm 4.3$ \\
\hline
$ {\rm [M/H]} = 0.0$, free $C_k$ & 750~K & 5.0 & $16.6 \pm 0.7$ & 750~K & 4.0 & $33.1 \pm 2.7$ \\
${\rm [M/H]} = +0.3$, free $C_k$ & 700~K & 5.0 & $19.8 \pm 1.2$ & 700~K & 4.0 & $27.7 \pm 1.7$ \\
\hline
\end{tabular}
\caption{Summary of the best fit model spectra and their associated $G$ values.}
\label{tab:bestfits}
\end{table*}

The best fitting model spectra in both the free scaling fits and the
constrained $R/D$ fits suggest properties that are consistent with
those estimated from our empirically determined luminosity for
Ross~458C.
That the best fits are all provided by the Solar metallicity BT Settl
models appears inconsistent with the [M/H] = +0.2--+0.3 estimates
for the system from \citet{burgasser2010}.
However, the $M_K$/metallicity and $V-K$/metallicity
relations of \citet{johnson2009} and \citet{schlaufman2010} that were
applied in that work are based on calibration from a relatively small
number of M~dwarfs, and Ross~458A lies at the early spectral type
extreme of the sample.
Additionally, the calibration samples are likely dominated by systems with ages
typically larger than 3~Gyr. Such systems show little evolution in the
$(V-K_s) - M_K$ plane with age, and thus any spread about the main
sequence may be attributed to metallicity, however the same can not be said
for dwarfs with the young age of Ross~458.
\citet{burgasser2010} also noted that the young age of Ross~458 may introduce a bias to higher metallicity, and \citet{morales2008} found systematically higher values of [Fe/H] for active stars when using the \citet{bonfils05} relations.

\subsection{Dust in the atmosphere of Ross 458C}

\begin{figure}
\includegraphics[height=250pt, angle=90]{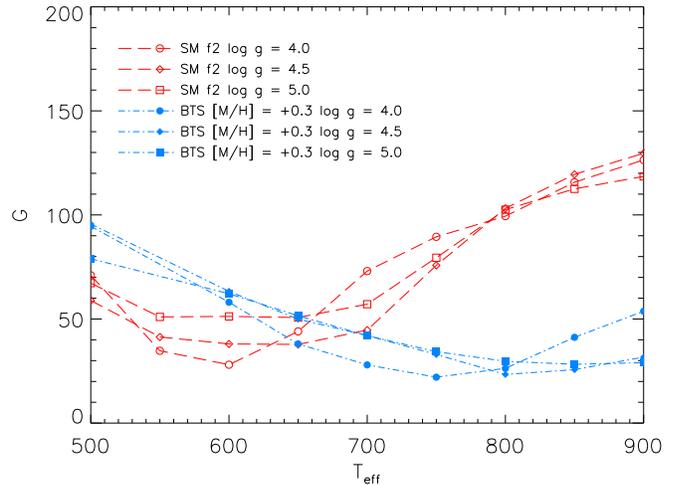}
\caption{The goodness-of-fit statistic, $G$, plotted for various
  models, considering just the near-infrared portion of the spectra
  with the value of the scaling factor $C$ evaluated such that 
  $G$ is minimised in each case \citep[see ][]{cushing08}. } 
\label{fig:dust}
\end{figure}

In Figure~\ref{fig:dust} we have plotted the
goodness-of-fit statistic, $G$, minimised for the case of a free scale factor, $C$, for Saumon \& Marley models in which condensate clouds are present in the photosphere and for BT Settl models.
The Saumon \& Marley atmospheres use the \citet{ackerman2001} cloud
model, which parameterises the efficiency of sedimentation of
condensate particles through an efficiency factor $f_{sed}$. 
Larger values represent faster particle growth and larger mean grain
sizes. Larger dust grains rain out of the atmosphere giving rise to
optically thin clouds. Smaller values of $f_{sed}$ thus correspond to thicker clouds.

Since no  Saumon \& Marley models with dust clouds are currently available 
for $\log K_{zz} = 6$, we have only fitted the near infrared spectrum
since a lower value of $K_{zz}$ will result in a poor fit at
longer wavelengths which would tend to dominate the statistic.
We found that the best fit for the Saumon \& Marley models
was for a dust sedimentation efficiency parameter $f_{sed}$ = 2, Solar
metallicity, $T_{\rm eff} = 600$K and $\log g = 4.0$ ($G = 28.03 \pm 0.3$).
These properties are consistent with what was found by \citet{burgasser2010},  who found that Saumon \& Marley models which included condensate opacity were able to
provide a better fit to their near infrared spectra than those without, leading them to
conclude that condensate clouds are an important opacity source in very cool
young T dwarf atmospheres. 
Whilst most of the cloud lies beneath the photosphere in this model, 
upper layers of cloud contribute to the near-infrared opacity, with an optical depth of $\sim 0.8$ in the $J$ band peak, such that the cloud is thick enough to define the photosphere in this region.

We have found that the BT Settl model with [M/H] = +0.3, $T_{\rm
  eff} = 750$K and $\log g = 4.0$ provides a significantly better fit
($G = 22.05 \pm 0.3$) than the dusty Saumon \& Marley models in the restricted near-infrared fits examined in this Section.
In the BT Settl models the main part of the cloud deck also lies well below the photosphere, but the upper cloud layers appear to have a smaller impact than in the case of the Saumon \& Marley models. 
This small amount of cloud opacity may account for the better match of the BT Settl models in the $J$~band peak compared to the cloud-free Saumon \& Marley models (see Figure~\ref{fig:speccomp}).
We can thus conclude that whilst condensates contribute some near-infrared opacity at the low $T_{\rm eff}$ of cool T dwarfs, the use of models with clouds thinner in vertical extent than the $f_{sed} = 2$ case is preferred. 
However, it should be noted that the difference in the dust opacity between the two models appears to be  relatively modest. 
Whilst $f_{sed} = 2$ corresponds to very dusty atmospheres at the $T_{\rm eff}$ of L~dwarfs, it results in a roughly similar dust signature to the BT Settl models at $T_{\rm eff} = 700$K.

We note, also, that the best fitting BT Settl models are
somewhat warmer than is found when the IRAC photometry is included in the
fit. This is likely due to the increased importance of the $K$ band
flux in the fit, which is consistently underestimated by the cool BT Settl
models.

\section{Comparison to the wider low-temperature sample}
\label{sec:scomp}

\begin{figure}
\includegraphics[height=270pt, angle=-90]{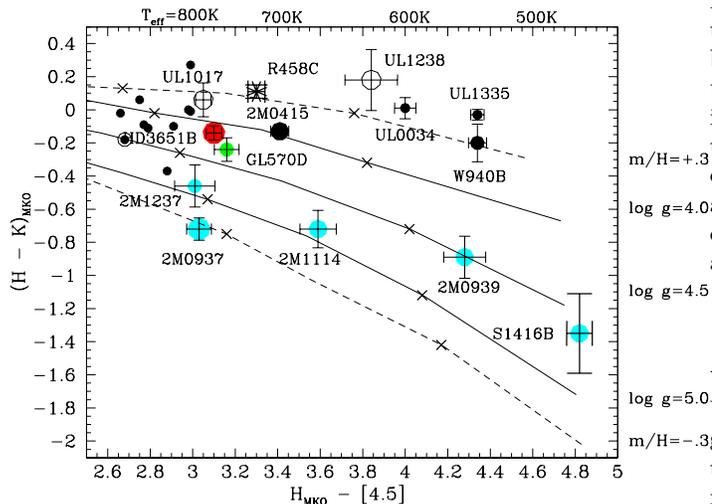}
\caption{Near- to mid-infrared colours of cool T~dwarfs compared to those of the Saumon \& Marley model colours, after Figure~10 of \citet{sandy10}. } 
\label{fig:h45plot}
\end{figure}

To place Ross~458C in context with the wider sample of very cool
T~dwarfs we have reproduced the
$H-K$/$H-{\rm [4.5]}$ colour-colour plot of \citet{sandy10} in Figure~\ref{fig:h45plot}.
Although the behaviour of the $H-[4.5]$ colour is likely poorly
understood (see Section \ref{subsec:models}) and the  models are known
to produce $H-K$ colours that are too blue, such a plot can still
provide useful insights to the relative properties of the cool dwarf
sample. 
The young age of Ross~458C and its location on this colour-colour plot
would appear to support the assertion that the low-temperature
T~dwarfs identified in UKIDSS to date are dominated by young, low-mass
objects \citep{sandy10}.
This is interesting in the context of the dearth of late T dwarfs
found in the Solar neighbourhood compared to what would be expected
given the mass function seen in young cluster and associations
\citep{ben10b}, although it is not yet clear if this is due to some
effect of the birthrate for brown dwarfs,  an as yet unidentified
selection effect or some problem with the evolutionary models.

\section{Summary}
\label{sec:summ}
We have presented near-infrared spectroscopy, photometry and warm-{\it
  Spitzer} IRAC photometry of the very cool brown dwarf Ross~458C.   
We assign it the spectral type T8.5p, where the `p' indicates the
peculiarity due to an enhanced $K$ band flux, presumably reflecting
the low-gravity associated with such a young object. 
We have used both BT~Settl and Saumon \& Marley model grids to
estimate the flux in regions not covered by our near-infrared
spectrum and IRAC photometry, and by using age constraints placed on
Ross~458AB by \citet{goldman10} we use the bolometric luminosity to
estimate that $T_{\rm eff} = 695 \pm 60$K and $\log g = 4.0-4.7$. 

We have also fitted various model spectra to the data and have found
that the Solar metallicity BT Settl model grid provide the best match to
the data for our derived parameter set, although both model sets provide similar estimates for
$T_{\rm eff}$ and gravity when used to fit for the parameters. The fitted
estimates are also consistent with our determination from the
bolometric flux.

We note, however, that these similar parameter estimates are despite
significant differences in the predicted spectrum in the 4--5$\micron$
region due to differences in approach to CO--CO$_2$ chemistry. 
The BT Settl spectra best reproduce the IRAC photometry, suggesting that chemical non-equilibrium effects are important in setting the CO-CO$_2$ ratio, although spectroscopy would be required to confirm this assertion.
The FLITECAM instrument \citep{flitecam} on board the Stratospheric
Observatory for Infrared Astronomy \citep[SOFIA;][]{sofia} may provide
the capability to distinguish the model sets for bright
objects such as Ross~458C.  For fainter objects, this spectral
region will be restricted to photometric data until the launch of the
James Webb Space Telescope. 

We have also examined the suggestion by \citet{burgasser2010} that
condensate clouds are an important source of opacity in the emergent
spectrum of Ross~458C. 
We find that the BT Settl models provide a better fit to the near infrared spectrum than the dusty models favoured by the \citet{burgasser2010} fits.
Although in both sets of models the main cloud deck lies well below the photosphere, in the Saumon \& Marley models the upper layers still effectively define the photosphere in near infrared spectral regions such as the $J$ band peak, whilst in the BT Settl case their influence is much smaller.
We thus conclude that condensate opacity does indeed appear to affect the near-infrared spectra of cool T dwarfs, although clouds with less vertical extent than is seen for the $f_{sed} = 2$ are preferred.

\section*{Acknowledgements}

We thank our anonymous referee for comments that have greatly improved the quality of the manuscript.   The authors wish to recognise and acknowledge the very significant cultural role and reverence that the summit of Mauna Kea has always had within the indigenous Hawaiian community.  We are most fortunate to have the opportunity to conduct observations from this mountain.
SKL is supported by the Gemini Observatory, which is operated by AURA,
on behalf of the international Gemini partnership of Argentina,
Australia, Brazil, Canada, Chile, the United Kingdom, and the United
States of America. 
CGT is supported by ARC grant DP0774000.
This research has made use of the SIMBAD database,
operated at CDS, Strasbourg, France, and has benefited from the SpeX
Prism Spectral Libraries, maintained by Adam Burgasser at
http://www.browndwarfs.org/spexprism.

\bibliographystyle{mn2e}
\bibliography{refs}

\end{document}